\documentclass{elsart}

\usepackage{amssymb,graphicx,amsmath,latexsym}
\newtheorem{theorem}{Theorem}
\newtheorem{definition}{Definition}
\newtheorem{corollary}{Corollary}

\def\eop{\hfill$\Box$}
\def\proof{{\em Proof: }}
\def\cG{\mathcal{G}}

\def\C{{\mathbb C}}

\begin{document}

\begin{frontmatter}



\title{Conditions for separability in generalized 
Laplacian matrices and nonnegative matrices as density matrices}


\author{Chai Wah Wu}
\ead{cwwu@us.ibm.com,chaiwahwu@member.ams.org}
\address{IBM Research Division,
Thomas J. Watson Research Center\\
P. O. Box 218,
Yorktown Heights, NY 10598, U. S. A.}

\begin{abstract}
Recently, Laplacian matrices of graphs are studied as
density matrices in quantum mechanics.  We continue this study and
give conditions for separability of generalized Laplacian matrices of weighted
graphs with unit trace.  In particular, we show that the Peres-Horodecki
positive partial transpose
separability condition is necessary and sufficient for separability in
${\mathbb C}^2\otimes {\mathbb C}^q$.  In addition, we present a
sufficient condition for separability of generalized Laplacian matrices 
and diagonally dominant nonnegative matrices.
\end{abstract}

\end{frontmatter}

\section{Introduction\label{sec:introduction}}
Due to novel applications of quantum mechanics in recent years such as
quantum teleportation, quantum cryptography and quantum computing
\cite{nielsen-quantum-2002}, there is much recent interest in
studying entanglement in quantum systems.  One important problem is to
determine whether a given state operator is entangled or not.  This is
especially difficult for mixed state operators.  In
Ref. \cite{braunstein:laplacian:2004}, normalized Laplacian matrices
of graphs are considered as density matrices, and their entanglement
properties are studied.  The reason for studying this subclass of
density matrices is that simpler and stronger conditions for
entanglements can be found.  In
Ref. \cite{braunstein:laplacian_graph:2005} a conjecture was proposed
on a necessary and sufficient condition for separability of such
density matrices and the conjecture was verified for some special
classes of graphs.

The purpose of this paper is to further this study and 
give some generalizations of these
results.  In particular, we show that the Peres-Horodecki positive partial transpose
condition is necessary
and sufficient for Laplacian matrices of weighted graphs to be
separable in ${\mathbb C}^2\otimes {\mathbb C}^q$.  
Furthermore, we give a sufficient condition for Laplacian matrices and
diagonally dominant nonnegative matrices
to be separable in $\C^p\otimes \C^q$.

\section{Density matrices, separability, and partial transpose}
We use $I$ and ${\bf 0}$ to denote the identity matrix and the zero
matrix respectively.  A state of a finite dimensional quantum
mechanical system is described by a state operator or a density matrix
$\rho$ acting on $\C^n$ which is Hermitian and positive semidefinite
with unit trace.  A state operator is called a pure state if 
it has rank one.  Otherwise the state operator is mixed.  An $n$ by
$n$ density matrix $\rho$ is separable in $\C^p\otimes \C^q$ with
$n=pq$ if it can be written as $\sum_{i} c_i \rho_i \otimes
\eta_i$ where $\rho_i$ are $p$ by $p$ density matrices and $\eta_i$
are $q$ by $q$ density matrices with $\sum_i c_i = 1$ and $c_i\geq
0$.\footnote{This definition can be extended to composite systems of
multiple states, but here we only consider decomposition into the
tensor product of two component states.}  A density matrix that is not
separable is called entangled.  Entangled states are necessary to
invoke behavior that can not be explained using classical physics and
enable applications such as quantum teleportation and quantum
cryptography.

We denote the $(i,j)$-th element of a matrix $A$ as $A_{ij}$.
Let $f$ be the canonical bijection between 
$\{1,\dots, p\}\times \{1,\dots,q\}$ and $\{1,\dots, pq\}$:
$f(i,j)  = (i-1)q+j$.  For a $pq$ by $pq$ matrix $A$,
if $f(i,j) = k$ and $f(i_2,j_2) = l$, we will sometimes write
$A_{kl}$ as $A_{(i,j)(i_2,j_2)}$.
\begin{definition}
The $(p,q)$-partial transpose $A^{pT}$ of an $n$ by $n$ matrix $A$,
where $n=pq$, is given by:
\[ A_{(i,j)(k,l)}^{pT} = A_{(i,l)(k,j)} \]
\end{definition}
We will usually remove the prefix ``$(p,q)$'' if $p$ and $q$ are clear
from context.  In matrix form, the partial transpose is constructed as
follows.  If $A$ is decomposed into $p^2$ blocks:
\begin{equation}\label{eqn:A}
 A = \left(\begin{array}{cccc}
A^{1,1} & A^{1,2} &\cdots & A^{1,p} \\
A^{2,1} & A^{2,2} & \cdots & A^{2,p} \\
\vdots  & \vdots  &  & \vdots \\
A^{p,1} & A^{p,2} & \cdots      & A^{p,p} \end{array}\right)
\end{equation}
where each $A^{i,j}$ is a $q$ by $q$ matrix, then 
$A^{pT}$ is given by:
\begin{equation}\label{eqn:Apt}
 A^{pT} = \left(\begin{array}{cccc}
(A^{1,1})^T & (A^{1,2})^T &\cdots & (A^{1,p})^T \\
(A^{2,1})^T & (A^{2,2})^T & \cdots & (A^{2,p})^T \\
\vdots  & \vdots & & \vdots \\
(A^{p,1})^T & (A^{p,2})^T & \cdots      & (A^{p,p})^T \end{array}\right)
\end{equation}

It is clear that if $A$ is Hermitian, then so is $A^{pT}$.
Peres \cite{peres:separability:1996} introduced the following necessary condition for
separability:
\begin{theorem} \label{thm:peres}
If a density matrix $\rho$ is separable, 
then $\rho^{pT}$ is positive semidefinite, i.e.
$\rho^{pT}$ is a density matrix.
\end{theorem}
Horodecki et al. \cite{horodecki:separability:1996} showed that this
condition is sufficient for separability in $\C^2\otimes \C^2$ and
$\C^2\otimes \C^3$, but not for other tensor products.  A density
matrix having a positive semidefinite partial transpose is often
referred to as the Peres-Horodecki condition for separability.

\section{Laplacian matrices of graphs as density matrices}
\begin{figure}[htbp]
\centerline{\includegraphics[width=1.2in,angle=-90]{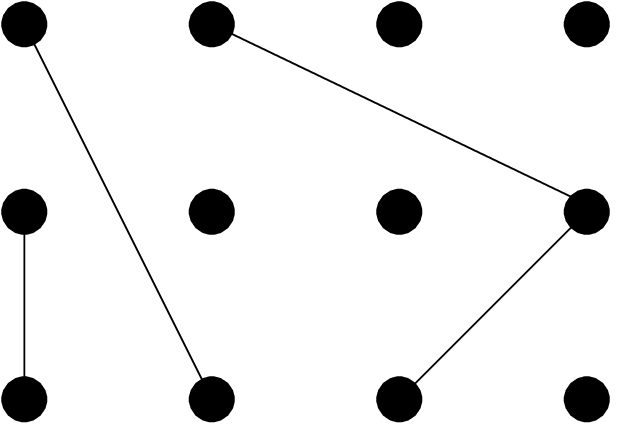}\hspace{1.1in}\includegraphics[width=1.2in,angle=-90]{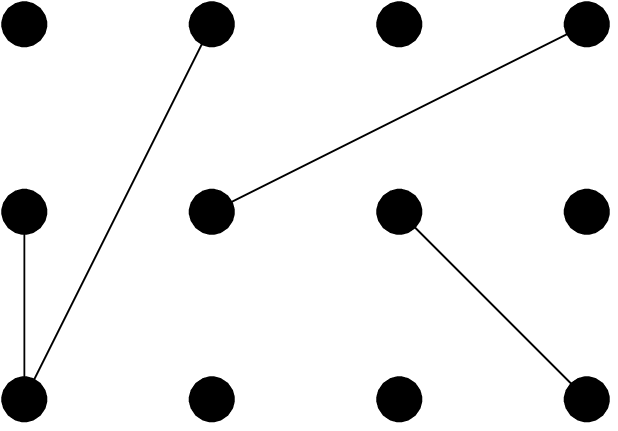}}
\centerline{(a)\hspace{2.6in}(b)}
\caption{(a) A graph $\cG$ and (b) its partial transpose $\cG^{pT}$ for the case $p=3$, $q=4$.}
\label{fig:graphs}
\end{figure}

For a graph $\cG$, the Laplacian matrix $L(\cG)$ is defined as $D-A$
where $D$ is the diagonal matrix of vertex degrees and $A$ is the
adjacency matrix.  The Laplacian matrix is symmetric positive
semidefinite, has zero row sums and has a simple zero eigenvalue if
and only if the graph is connected \cite{godsil:alg_graph:2001}.  Let
$|E|$ be the total number of edges in the graph.  Then
$\frac{1}{2|E|}L(\cG)$ has unit trace and is thus a density matrix.
In Ref. \cite{braunstein:laplacian:2004} properties of such density
matrices are obtained by studying the properties of the underlying
graph.
If $A$ is the adjacency matrix of $\cG$, let  $\cG^{pT}$ 
be the graph with adjacency matrix $A^{pT}$. 
Graphically, $\cG^{pT}$ is obtained from $\cG$ as follows.  Let the vertex $k = f(i,j)$ be located at coordinate $(i,j)$.  Then
$\cG^{pT}$ is obtained from $\cG$ by reflecting each edge vertically (or horizontally) around the midpoint of the edge.  Fig. \ref{fig:graphs} shows an example for $p=3$, $q=4$.

In Ref. \cite{braunstein:laplacian_graph:2005} it was shown that
the vertex degrees of $\cG$  equaling the vertex degrees of $\cG^{pT}$ is a necessary condition for separability and it was conjectured that this
 is also a sufficient condition. 
The sufficiency is shown for perfect matchings in
$\C^2\otimes \C^q$ and nearest point graphs.

\section{Generalized Laplacian matrices of weighted graphs}
Let $S$ be the set of symmetric real matrices with nonnegative row sums and
nonpositive off-diagonal elements.  Then $S$ is a subset of
generalized Laplacian matrices as defined in Ref. \cite{godsil:alg_graph:2001}.
We can associate a simple weighted graph to a matrix $A\in S$:
for $i\neq j$, 
$A_{ij}\neq 0$ corresponds to an edge from vertex $i$ to vertex $j$ with weight $-A_{ij} > 0$.  
Matrices in $S$ are positive semidefinite.  
If the corresponding graph is connected, 
then the smallest eigenvalue is simple.
Let $S_1$ be the matrices in $S$ with unit trace and let $S_1^0$ be matrices
in $S_1$ with zero row sums.  We focus on $S_1$ and $S_1^0$ which 
are sets of 
density matrices.  
We start with the following necessary condition for separability:

\begin{theorem}\label{thm:row_sums}
Let $A$ be a density matrix with zero row sums.  If $A^{pT}$ does not have zero row sums,
then $A$ is not separable.
\end{theorem}
\proof 
Let $e = (1,\dots 1)^T$.  Then $0 = e^TAe = \sum_{ij}A_{ij} = 
\sum_{ij}A^{pT}_{ij} = e^TA^{pT} e$.  Since $A^{pT}e\neq 0$, if $e$ is an eigenvector of $A^{pT}$, it means that it corresponds to a nonzero eigenvalue of $A^{pT}$ which contradicts the fact that $e^TA^{pT} e = 0$.  Therefore
$e = \sum_i a_iv_i$ is a linear combination of several eigenvectors $v_i$ of $A^{pT}$, one of which is not in the kernel of $A^{pT}$.  Since $0 =  e^TA^{pT} e
= \sum_i \lambda_i |a_i |^2 \|v_i\|^2$, this means that one of the $\lambda_i$ is negative, which implies that $A^{pT}$ is not positive semidefinite, hence by 
Theorem \ref{thm:peres}, $A$ is not separable.
\eop

Note that Theorem \ref{thm:row_sums} is applicable to general density
matrices, i.e. not necessarily Laplacian matrices of graphs.  The
above result generalizes Theorem 1 in
Ref. \cite{braunstein:laplacian_graph:2005} and is weaker than
the Peres-Horodecki condition, but is easier to verify when $A$ happens
to have zero row sums.
On the other hand, for matrices in $S_1^0$, this condition is equivalent to the Peres-Horodecki
condition.

\begin{theorem}  \label{thm:peres_equivalent}
Let $A$ be a density matrix in $S_1^0$. The matrix $A^{pT}$ has zero row sums
if and only if $A^{pT}$ is positive semidefinite.
\end{theorem}
\proof One direction follows from the proof of Theorem
\ref{thm:row_sums}.  If $A^{pT}$ has zero row sums, it is a matrix in
$S_1^0$ and hence positive semidefinite.  \eop

It is easy to show that for $A\in S$ with corresponding graph $\cG$, 
$A^{pT}$ having the same row sums as $A$
is equivalent to the vertex degrees of $\cG$ and $\cG^{pT}$ being equal.\footnote{The vertex degree of $v$ in a weighted graph is the sum of the weights of all the edges connected to $v$.}

For $A\in S$, an 
edge corresponding to $A_{(i,j)(i',j')}\neq 0$ is called entangled if
$i\neq i'$ and $j\neq j'$.  It's easy to see that if an edge
corresponding to $A_{(i,j)(i',j')}$ is not entangled, then
$A_{(i,j)(i',j')} = A^{pT}_{(i,j)(i',j')}$. In
Ref. \cite{braunstein:laplacian:2004} it was conjectured that
normalized Laplacian matrices of graphs where all entangled edges are
adjacent to the same vertex are not separable.  By Theorem
\ref{thm:row_sums} this conjecture is true.  In particular, it is true
for the larger set of density matrices in $S_1^0$ since the degree of
this vertex must necessarily decrease in $A^{pT}$.

\section{Sufficient and necessary conditions for separability}\label{sec:sep_cond}

\begin{theorem}\label{thm:unitary}
Let $F$ be a $q$ by $q$ matrix of the form
$F = \left(\begin{array}{cc} U &  \\  & {\bf 0} \end{array}\right)$
and $D$ be a $q$ by $q$ diagonal matrix 
of the form 
$D  = \left(\begin{array}{cc} I & \\  & {\bf 0} \end{array}\right)$
where $U$ is an unitary matrix of the same size as $I$.
Then  the matrix
\[ A = \frac{1}{2{\mbox{Tr}}(D)}\left(\begin{array}{cc} D & F \\ F^{\dag} & D \end{array}\right)
\]
is a density matrix and is separable in $\C^2\otimes \C^q$.
\end{theorem}
\proof
Since $U$ is unitary, it can be written as $U = \sum_i \lambda_i
v_iv_i^{\dag}$ where $\lambda_i $ and $v_i$ are the eigenvalues and
eigenvectors of $U$ respectively and $v_i^{\dag}$ denotes the complex
conjugate transpose of $v_i$.  Furthermore, all eigenvalues of $U$ are on the
unit circle and written as $\lambda_i = e^{\theta_i}$ with imaginary
numbers $\theta_i$.  Define
\[ w_i = \left(\begin{array}{c}v_i \\
0\\\vdots \\ 0\end{array}\right)\]
This means that $F = \sum_i \lambda_i w_i
w_i^{\dag}$.  Furthermore, $\sum_i v_i v_i^{\dag} = I$ implies that
$\sum_i w_i w_i^{\dag} = D$.  Define $x_i =
\left(\begin{array}{c}e^{\frac{\theta_i}{2}} \\
e^{\frac{-\theta_i}{2}}\end{array}\right) \otimes w_i $.  Then
\[ \sum_i x_i x_i^{\dag} =
\left( \begin{array}{cc} \sum_i w_i w_i^{\dag} &  \sum_i \lambda_i w_i w_i^{\dag} \\
\sum_i \overline{\lambda_i} w_i w_i^{\dag} & \sum_i w_i w_i^{\dag} \end{array}\right) 
= \left(\begin{array}{cc} D & F \\ F^{\dag} & D \end{array}\right)
\]
which shows that $A$ is positive semidefinite and is separable.
\eop

\begin{definition}
$C$ is a simple circuit matrix if there exists distinct integers 
$i_1,\dots, i_k$ $(k\geq 1)$
such that $C_{i_mi_{m+1}} = 1$ for $m = 1,\dots, k-1$, $C_{i_ki_1} = 1$ 
and $C_{ij} = 0$ otherwise.
\end{definition}

\begin{corollary}\label{cor:circuit}
If $C$ is a $q$ by $q$ simple circuit matrix and $D$ is the diagonal matrix with the row sums of $C$ on the diagonal,
then  the matrix
\[ A = \frac{1}{2{\mbox{Tr}}(D)}\left(\begin{array}{cc} D & -C \\ -C^T & D \end{array}\right)
\]
is a density matrix and is separable in $\C^2\otimes \C^q$.
\end{corollary}
\proof
There exists 
permutation matrices  $P$, $Q$ such that $Q^TCQ$ is of the form 
$\left(\begin{array}{cc} P &  \\  & {\bf 0} \end{array}\right)$.
Therefore without loss of generality, 
we assume that $C$ is of the form
$C = \left(\begin{array}{cc} P &  \\  & {\bf 0} \end{array}\right)
$.
This means that 
$D  = \left(\begin{array}{cc} I & \\  & {\bf 0} \end{array}\right)$.
Since $-P$ is unitary, the result follows from Theorem \ref{thm:unitary}.
\eop

\begin{definition}
A matrix is line sum symmetric if the $i$-th column sum is equal to the $i$-th row sum for each $i$.
\end{definition}

\begin{theorem} \label{thm:separable_s1}
Let $A$ be a $2q$ by $2q$ density matrix in $S_1$.   If $A^{pT}$ has the same row sums
as $A$, then the matrix $A$ is separable in $\C^2\otimes \C^q$.
\end{theorem}
\proof 
It is easy to see that 
$A$ can be decomposed into:
\[ A = \left( \begin{array}{cc} A_1 &  \\  & {\bf 0} \end{array}\right) 
+ \left( \begin{array}{cc} {\bf 0} &  \\  & A_2 \end{array}\right)
+  \left( \begin{array}{cc} D_1 & -B \\ -B^T & D_2 \end{array}\right)
\]
where $A_1$ and $A_2$ are symmetric matrices and $D_1$ and $D_2$ are diagonal matrices with the row sums of $B$ and $B^T$ on the diagonal respectively.
It is easy to see that $A_1$ and $A_2$ are in $S$ and thus are 
positive semidefinite.  The first term  $\left( \begin{array}{cc} A_1 &  \\  & 
{\bf 0} \end{array}\right)$ can be separated as
$ \left( \begin{array}{cc} 1 & 0 \\ 0 & 0 \end{array}\right) \otimes A_1 $.
Similarly the second term can be separated as well.  As for the third term, 
the matrix $B$ is
a nonnegative matrix.  Suppose $A^{pT}$ has the same row sums as $A$.
This implies that the $i$th column sum of $B$ is equal to its $i$th
row sum. This means that $B$ is line sum symmetric.  Using results in
network flow theory, it was shown in
Ref. \cite{dantzig:line-sum-symmetric:1985} that $B$ can be written as
$\sum_i \alpha_i C_i$, where $C_i$ are simple circuit matrices and
$\alpha_i \geq 0$.  This together with Corollary \ref{cor:circuit}
implies that $A$ is separable in $\C^2\otimes \C^q$.\eop

Our main result shows that for density matrices in $S_1^0$, the Peres-Horodecki
condition (which by Theorem \ref{thm:peres_equivalent} is equivalent
to $A^{pT}$ having zero row sums) is sufficient and necessary for
separability in $\C^2\otimes \C^q$.

\begin{theorem}
Let $A$ be a $2q$ by $2q$ density matrix in $S_1^0$.  The matrix $A$ is separable in $\C^2\otimes \C^q$ if and only if $A^{pT}$ has zero row sums.
\end{theorem}
\proof Follows from Theorems \ref{thm:row_sums} and \ref{thm:separable_s1}.\eop

\section{A sufficient condition for separability}
The techniques in Section \ref{sec:sep_cond} can be applied to density
matrices in $S_1$ acting on the more general tensor product $\C^p \otimes
\C^q$.  In fact, the same arguments can be used to prove the following sufficient condition for separability:
\begin{theorem}\label{thm:sepgeneral}
If an $n$ by $n$ density matrix $A$ in $S_1$ written in the form
Eq. (\ref{eqn:A}) is such that the matrices $A^{i,j}$ are line sum
symmetric, then $A$ is separable in $\C^p\otimes \C^q$.
\end{theorem}

\begin{corollary}\label{cor:local}
Let $A$ be a density matrix in $S_1$ be such that $A_{(i,j)(i',j')}\neq 0$
implies that $|i-i'|\leq 1$.\footnote{Graphically, if the vertices are arranged as in Fig. \ref{fig:graphs}, then the edges connect vertices in the same row or in adjacent rows.}  
If $A^{pT}$ has the same row sums as $A$, then
$A$ is separable in $\C^p\otimes \C^q$.
\end{corollary}
\proof  
By hypothesis, $A$ is in block tridiagonal form:
\begin{equation}\label{eqn:Atridiagonal}
 A = \left(\begin{array}{cccc}
A^{1,1} & A^{1,2} & {\bf 0} &  \\
A^{2,1}  & A^{2,2} & A^{2,3} & \ddots\\
& \ddots & \ddots & \ddots \\
& {\bf 0} &   A^{p,p-1}   & A^{p,p} \end{array}\right)
\end{equation}
If $A^{pT}$ has the same row sums as $A$, 
then $A^{1,2}$ is line sum symmetric which in turns means that $A^{2,1} = (A^{1,2})^T$ is line sum symmetric.  This implies that $A^{2,3}$ is line sum symmetric etc.  The result then follows from Theorem \ref{thm:sepgeneral}.\eop

\begin{corollary}\label{cor:local2}
Let $A$ be a density matrix in $S_1^0$ be such that $A_{(i,j)(i',j')}\neq 0$
implies that $|i-i'|\leq 1$.  Then $A$ is separable in $\C^p\otimes \C^q$ 
if and only if 
$A^{pT}$ has zero row sums.
\end{corollary}
\proof Follows from Theorem \ref{thm:row_sums} and Corollary \ref{cor:local}.
\eop

For a normalized Laplacian matrix $A$ of a graph such that
$A_{(i,j)(i',j')} \neq 0$ implies $|i-i'|\leq 1$ and $|j-j'|\leq
1$ (the so-called nearest point graphs), it was shown in Ref. \cite{braunstein:laplacian_graph:2005} via a
combinatorial argument that $A$ is separable if and only if $A^{pT}$
has zero row sums.  Corollary \ref{cor:local2} is a generalization of
this result as the condition $|j-j'|\leq 1$ is not necessary.

\section{Diagonally dominant nonnegative symmetric matrices}
Let $V$ be the set of diagonally dominant nonnegative symmetric matrices, i.e.
nonnegative symmetric matrices such that $A_{ii} \geq \sum_{i\neq j}A_{ij}$ for
all $i$.  By Gershgorin's circle criterion \cite{marcus:matrices:1992} 
matrices in $V$
are positive semidefinite.
Let $V_1$ be the set of matrices in $V$ with unit trace.  Then $V_1$ are density matrices.
We obtain results for $V_1$ which are analogous to those for $S_1$.
For instance, we have the following Corollary to Theorem \ref{thm:unitary}:
\begin{corollary}\label{cor:circuit_V1}
If $C$ is a $q$ by $q$ simple circuit matrix and $D$ is the diagonal matrix with the row sums of $C$ on the diagonal,
then  the matrix
\[ A = \frac{1}{2{\mbox{Tr}}(D)}\left(\begin{array}{cc} D & C \\ C^T & D \end{array}\right)
\]
is a density matrix and is separable in $\C^2\otimes \C^q$.
\end{corollary}

Corollary \ref{cor:circuit_V1} and the  argument in Section \ref{sec:sep_cond}
are used to prove the following:

\begin{theorem}
Let $A$ be a $2q$ by $2q$ density matrix in $V_1$.  If $A^{pT}$ has the same row sums as $A$, then the matrix $A$ is separable in $\C^2\otimes \C^q$.
\end{theorem}

\begin{corollary}\label{cor:local_V1}
Let $A$ be a density matrix in $V_1$ be such that $A_{(i,j)(i',j')}\neq 0$
implies that $|i-i'|\leq 1$.
If $A^{pT}$ has the same row sums as $A$, then
$A$ is separable in $\C^p\otimes \C^q$.
\end{corollary}

\begin{theorem}\label{thm:sepgeneral_V1}
If an $n$ by $n$ density matrix $A$ in $V_1$ written in the form
Eq. (\ref{eqn:A}) is such that the matrices $A^{i,j}$ are line sum
symmetric, then $A$ is separable in $\C^p\otimes \C^q$.
\end{theorem}

\section{Conclusions}
We study separability criteria for density matrices that can be
expressed as generalized Laplacian matrices of weighted graphs or
diagonally dominant nonnegative matrices.  In particular, we show that
the well-known Peres-Horodecki necessary condition for separability is
in fact sufficient for Laplacian matrices in $\C^2\otimes \C^q$.

\bibliography{chua_ckt2,chaos,secure,synch,misc,stability,cml,algebraic_graph,graph_theory,control,optimization,adaptive,top_conjugacy,ckt_theory,cnn2,matrices,chaos_comm,markov,quantum}
\end{document}